# Critical behavior and magnetic relaxation dynamics of $Nd_{0.4}Sr_{0.6}MnO_3$ nanoparticles

S. Kundu[1], and T. K. Nath[2]

[1,2]Department of Physics and Meteorology, Indian Institute of Technology, Kharagpur 721302, India

[1]Rajiv Gandhi University of Knowledge Technologies, Nuzvid 521201, Andhra Pradesh, India

[2]Email: tnath@phy.iitkgp.ernet.in

## Abstract

*Detailed dc and ac magnetic properties of chemically synthesized $Nd_{0.4}Sr_{0.6}MnO_3$ with different particle size (down to 27 nm) have been studied in details. We have found ferromagnetic state in the nanoparticles, whereas, the bulk $Nd_{0.4}Sr_{0.6}MnO_3$ is known to be an A-type antiferromagnet. A Griffiths-like phase has also been identified in the nanoparticles. Further, critical behavior of the nanoparticles has been studied around the second order ferromagnetic-paramagnetic transition region ($|(T-T_C)/T_C| \leq 0.04$) in terms of modified Arrott plot, Kouvel-Fisher plot and critical isotherm analysis. The estimated critical exponents ($\beta, \gamma, \delta$) are quite different from those predicted according to three-dimensional mean-field, Heisenberg and Ising models. This signifies a quite unusual nature of the size-induced ferromagnetic state in $Nd_{0.4}Sr_{0.6}MnO_3$. The nanoparticles are found to be interacting and do not behave like ideal superparamagnet. Interestingly, we find spin glass like slow relaxation of magnetization, aging and memory effect in the nanometric samples. These phenomena have been attributed to very broad distribution of relaxation time as well as to inter-particle interaction. Experimentally, we have found out that the dynamics of the nanoparticle systems can be best described by hierarchical model of spin glasses.*

---

[1]Corresponding author: souravphy@gmail.com

Tel: +91-3222-283862
Area code: 721302,
INDIA

**1. Introduction**

In recent times, the study of the properties of nanoparticles of antiferromagnetic oxides including the strongly correlated manganites has been a topmost research interest. It has been found in many occasions that the transition metal antiferromagnetic oxides display many novel properties when they are taken in nanometric form. For instance, CuO nanoparticles show a drastic change in the Neel temperature compared to the bulk counterpart [1]. A similar phenomenon was observed later on in the system of $Cr_2O_3$ nanoparticles by D. Tobia et al. [2]. Additionally, they have observed that the spin flop field decreases with the decrease of particle size. The change in the magnetic properties of NiO nanoparticles was attributed to the distribution of vacancies around the surface region within a core-shell type model [3]. In some recent other studies the observed properties in the nanoparticles of different oxides (CoO, Co/CoO, MnO etc.) has been attributed to the so called core - shell structure [4-6].

Similar studies on the effect of size reduction of antiferomagnetic manganites have also been carried out. In case of a perovskite type manganite $CaMnO_{3-\delta}$ it was found that the reduction of size led to the observation of the phenomena of exchange bias effect [7]. However, as far as antiferromagnetic (AFM) manganites are concerned, most of the studies regarding nanometric effect have been performed on half-doped manganites which display charge-orbital order [8-11]. On size reduction, the AFM state of the half-doped manganites is generally found to destabilize with a concomitant appearance of weak ferromagnetism. Effects of surface disorder, surface pressure, strain modification etc. have been attributed to such observation. At this stage, it appears that the study of the reduction of particle size in case of AFM manganites other than the family having charge -orbital order is very important.

Moreover, in any case, it is necessary to investigate the nature of the size-induced magnetic state in the nanoparticles.

In this context, the study of the size effect on $Nd_{0.4}Sr_{0.6}MnO_3$ which have a antiferromagnetic ground state in bulk form with A-type spin order and is fairly away from the charge ordered region ($Mn^{3+} : Mn^{4+} \sim 0.5$) [12] is relevant. A detailed characterization of its magnetic properties has been carried out with the change of particle size down to 27 nm.

## 2. Experimental details

The samples of $Nd_{0.4}Sr_{0.6}MnO_3$ have been synthesized through chemical "pyrophoric" reaction route [13]. Stoichiometric amount of high purity $Nd_2O_3$, $Sr(NO_3)_2$ and $Mn(CH_3COO)_2$ were dissolved in distilled water with proper amount of $HNO_3$. Then triethanolamine (TEA) is added in the solution with 4:1:1 molar ratio with the metal ions (Nd,Sr:Mn:TEA=1:1:4). The finally obtained solution was then heated at 180 $^0$C with continuous stirring. Ultimately, combustion took place and a black fluffy powder was obtained. This as-prepared powder was divided into four parts, ground thoroughly and calcinated at different temperatures (600 $^0$C to 1150 $^0$C) in air for 5 hours to obtain single phase samples of $Nd_{0.4}Sr_{0.6}MnO_3$ with different particle size. Since all the samples are derived from the same as prepared powder, any additional change in stoichiometry in the samples is not expected to take place. The samples are designated in this manuscript as S600, S750, S850, S1150, respectively, after their calcination temperature.

Structural characterization of the samples has been carried out through high resolution x-ray diffraction (HRXRD) with Cu - $K_\alpha$ radiation ($\lambda = 1.5418$ Å), transmission electron microscopy (TEM) and field emission scanning electron microscopy (FESEM). The low field magnetic measurements have been carried out employing homemade low field vibrating sample magnetometer [14] and a homemade ac susceptometer. A high precision, digital signal processing lock – in – amplifier (Stanford Research Systems, SR830) was utilized to detect the signal. The temperature was monitored and controlled by employing a precision PID temperature controller (Lakeshore, 331S) with better than ±50 mK stability. The high field magnetic measurements have been carried out employing a VSM-SQUID (Quantum Design).

## 3. Results

### 3.1. Structural characterization

The x- ray diffraction data (high resolution) shown in the Fig. 1(a) and (b) for S1150 and S600 samples, respectively, display polycrystalline nature of the samples and broadening of the peaks with the decrease of calcination temperature signifying the decrease of average crystallite size. All the x-ray spectra have been analysed through Rietveld refinement using Maud programme [15]. No impurity phase has been found. The x-ray data have been fitted well with *I4/mcm* space group of tetragonal crystal structure and the obtained information are summarized in Table 1. We see that the structural change is negligibly small with the change of particle size. The TEM images of S600 sample in Fig. 1(c) and the plot of size distribution of the particles in the Fig. 1(d) show that the average grain size of S600 sample is about 27 nm. Similarly, the average grain size of S750 sample is found to be about 45 nm as evident from TEM image in Fig. 1(e) and the plot of particle size distribution shown in     Fig. 1 (f). From the FESEM images of the samples S850 and S1150 shown in the Fig. 1(g) and Fig. 1 (h), the average particle size is found to be about 95 nm and 300 nm, respectively.

### 3.2. Dc magnetization study: size-induced ferromagnetism and critical behavior

The magnetization of all the samples has been measured with respect to temperature during heating after cooling the sample without (ZFC) and with field (FC). Figure 2 shows the ZFC and FC magnetization data of all the samples measured at 50 Oe. The ZFC magnetization of S1150 sample (Inset of Fig. 2) increases with temperature, exhibiting a peak around 225 K, and decreases thereafter. This nature is quite similar to an ordinary AFM – PM (paramagnetic) transition. The value of the peak-temperature obtained in bulk-like S1150 is in good agreement with the bulk phase diagram of Nd-Sr-Mn-O system showing an A-type AFM to PM transition at around the same temperature [12]. This sample exhibits bifurcation between the ZFC and FC curves. Such a bifurcation is very unlikely in an ordinary antiferromagnetic material. However, we attribute this bifurcation to the uncompensated spins at the grain boundary region of this polycrystalline material (S1150). During field cooling these spins will freeze and point to the field direction. This gives rise to an observable bifurcation between ZFC-FC magnetizations. On reduction of particle size the peak in the ZFC curve of the S850, S750 and S600 samples shows a broad peak with a concomitant enhancement of the magnetization value compared to that of the S1150 sample. For these three samples an enhanced bifurcation between the FC and ZFC curve is also found. The overall M(T) behavior of the nanometric samples (S850, S750 and S600) is quite similar to those found in magnetic nanoparticle systems like superparamagnets. This behavior also indicates that the AFM order in the bulk sample has been modified due to size reduction. At this stage it is necessary to inspect the nature of magnetic state in the nanoparticles.

We have carried out a detailed dc magnetization study in one of the nanometric systems, namely the S750 sample. Firstly, we have measured the magnetization of this sample as a function of magnetic field at different temperatures. The M-H curve at 10 K is shown in the Fig. 3(a).This curve shows a clear hysteresis with a qualitatively ferromagnetic (FM) nature. The magnetization does not saturate within the applied field range. From the Arrott plot [16] at 10 K as shown in the inset of Fig. 3(a), a positive intercept is found with a spontaneous magnetization ($M_s$) of about = 18.8 emu/g. This confirms the presence of a size-induced ferromagnetic (FM) ground state in the S750 sample. The non-saturating M-H nature is possibly due to the presence of some fraction of AFM phase in the sample. Furthermore, the plot of inverse susceptibility ($1/\chi$) as a function of temperature is found to be linear in the high temperature paramagnetic regime showing Curie-Weiss type behavior. Employing the equation,

$$\chi = C/(T - \theta_{CW}),\qquad(1)$$

the value of $\theta_{CW}$ is found to be equal to 225 K (Fig. 3(b)). Importantly, the positive value of Curie-Weiss temperature further confirms the presence of FM exchange interaction in this sample (S750). The calculated effective PM moment ($\mu_{eff} = 2.82\sqrt{C}$, C is the Curie-Weiss constant) is found to be 4.82$\mu_B$ which is nearly equal to the theoretically calculated value of $\sqrt{0.4 \times \mu_{Nd^{3+}}^2 + 0.4 \times \mu_{Mn^{3+}}^2 + 0.6 \times \mu_{Mn^{4+}}^2}$ = 4.87 $\mu_B$.

As the temperature is lowered, evidently, $\chi^{-1}$ deviates from Curie-Weiss behavior in the form of a down-turn (Fig. 3(b)). This down-turn is a characteristic feature of Griffiths phase (GP) [17] which is a recent issue of some interest regarding the physics of manganites. Observation of a Griffiths like phase due to reduction of grain size seems quite interesting. The concept of Griffiths phase is originally proposed for a diluted Ising ferromagnet with only a fraction of sites occupied by the magnetic moments [18,19]. In a temperature regime, between the magnetic long range ordering temperature of the actual system ($T_C$) and that of the undiluted system ($T_G$), the system is said to be in Griffiths phase. In real systems like manganites this phase (in the temperature regime $T_C < T < T_G$) is manifested as a magnetically inhomogenous cluster like phase where no long range magnetic order is established but short range correlated embedded FM clusters exists in paramagnetic (PM) matrix [17-22]. Below $T_G$ susceptibility deviates from a Curie-Weiss type behaviour.

The temperature where the onset of down turn is observed is known as the Griffiths temperature $T_G$. In case of S750 sample we find that $T_G = 290$ K ($\pm 10$ K) as shown in the Fig. 3(b). One important fact about this Griffiths like phase is that in the Griffiths regime, namely in the regime $T_C < T < T_G$ no spontaneous magnetization exists (as observed from Fig. 4) due to the absence of any long range static FM order.

Usually, the Griffiths singularity is characterized by the exponent $\lambda (0 \leq \lambda \leq 1)$ obtained from the relation $1/\chi \propto (T - T_C^R)^{1-\lambda}$ [17]. The parameter $T_C^R$ is the critical temperature of the random ferromagnet where susceptibility diverges. According to Bray's generalization [19] of bond distribution of the concept proposed by Griffiths, the GP is supposed to be present in the regime $T_C^R < T < T_G$. In this temperature regime the singularity arises in the thermodynamic properties. Evidently, the exponent $\lambda$ determines the degree of deviation from the Curie-Weiss behaviour and hence the strength of the GP. The plot of $\chi^{-1}$ as a function of reduced temperature $(T/T_C^R - 1)$ in log scale is shown in the Fig. 3(c). The plot displays two distinct linear regimes. From the linear fit in GP regime, the obtained values of $T_C^R$ and $\lambda$ are found to be 225 K and 0.69, respectively. The temperature $T_C^R$ is chosen in such an away that $\lambda$ becomes nearly 0 in the PM regime. So, $T_C^R$ is actually equivalent to $\theta_{CW}$. The value of $\lambda$ is comparable to those values obtained for different manganites [17,20,21].

To have more insight into the size-induced FM state in S750 sample we have performed detailed scaling analysis around the critical regime (FM-PM transition regime). According to scaling hypothesis, for second order phase transition, the spontaneous magnetization $M_s$ below $T_C$, the inverse of the initial susceptibility $\chi_0^{-1}$ above $T_C$ and magnetization M at $T_C$ obey the following relations [22] :

$$M_s(T) = M_0(-\varepsilon)^\beta, \ \varepsilon < 0 \qquad (2)$$

$$\chi_0^{-1}(T) = \Gamma(\varepsilon)^\gamma, \ \varepsilon > 0 \qquad (3)$$

$$M = XH^{1/\delta} \qquad (4)$$

Here, $\varepsilon = (T-T_C)/T_C$ is the reduced temperature; $M_0$, $\Gamma$, X are the critical amplitudes and $\beta$, $\gamma$, $\delta$ are the critical exponents. From the Arrott plot ($M^2$ vs. H/M plot) over a wide range of temperature for S750 sample as shown in the Fig. 4(a), we find that the curves are not straight and parallel to each other in the high field regime. This indicates that the magnetization of this sample does not obey mean field type behavior ($\beta$=0.5, $\gamma$=1). Moreover, positive slope of the curves indicates the second order nature of the FM-PM phase transition in the sample [23]. So, we have taken use of the modified Arrott plot on the basis of the Arrott-Noakes equation of state given by [24],

$$\left(\frac{H}{M}\right)^{1/\gamma} = a\frac{(T-T_C)}{T} + bM^{1/\beta} \qquad (5)$$

to calculate the critical exponents associated with the phase transition. Here, a and b are constants. The modified Arrott plot ($M^{1/\beta}$ vs. $(H/M)^{1/\gamma}$) is shown in the Fig. 4 (b) for $|\varepsilon| \leq 0.04$. The isotherms around the critical regime are straight lines and parallel to each other in the high field regime. The proper choice of the exponents $\beta$ and $\gamma$ is done following the procedure reported earlier [25,26]. We find that $\beta$ and $\gamma$ are equal to 0.884 and 0.782, espectively. From Widom scaling relation we find that $\delta$=1+ $\gamma/\beta$ = 1.884. Extrapolation of the isotherm at 251 K passes through the origin; thus indicating that $T_C$ = 251 ± 1.5 K.

To determine the $T_C$ as well as the critical exponents, we adopted another method known as the Kouvel-Fisher (KF) method [27]. This method suggests that the plot of $M_S (dM_S/dT)^{-1}$ and $\chi_0^{-1} (d\chi_0^{-1}/dT)^{-1}$ as a function of T yield straight lines with slopes $1/\beta$ and $1/\gamma$, respectively. The intercepts of these lines on T-axis give the value of $T_C$. From the KF plot shown in the Fig. 5 (a), the estimated critical exponents are $\beta$ =0.933 and $\gamma$ = 0.810. The value of $T_C$ found from the intercepts by the fitted straight lines is 251.5 K. All these estimated parameters from KF plot are having <5% error in their values. We also find that the values of critical exponents and $T_C$ estimated from KF plot are within ~ 5% of those obtained from modified Arrott plot. Furthermore, from the slope of critical isotherm (M(H) at T= $T_C$ $\cong$251 K) plotted in log-log scale (inset of Fig. 5(a)), the estimated value of $\delta$ is found to be equal to 1.976. This value is also within ~5% of that estimated from Widom relation. This implies that the estimated critical exponents are reasonably accurate. Importantly, we observe that the estimated critical exponents of S750 sample are significantly different from the values predicted according to standard 3D Heisenberg ($\beta$=0.365, $\gamma$= 1.386, $\delta$= 4.80) or 3D Ising ($\beta$=0.325, $\gamma$=1.241, $\delta$=4.82) models.

Next we compare our data with the prediction of the scaling theory,

$$M/|\varepsilon|^{\beta} = f_{\pm}(H/|\varepsilon|^{(\beta+\gamma)}),\qquad(6)$$

where, '+' and '−' signs are for temperatures above and below $T_C$, respectively. Here, f is a regular function. This relation further predicts that $M/|\varepsilon|^{\beta}$ plotted as a function of $H/|\varepsilon|^{(\beta+\gamma)}$ will produce two different set of curves, one for temperatures below $T_C$ and the other for the temperatures above $T_C$. This plot (using the values of critical exponents and $T_C$ obtained from KF plot) shown in the Fig. 5 (b) clearly displays two different branches corresponding to data points for $T<T_C$ and $T>T_C$. This also signifies the reliability of the estimated values of the critical exponents.

Such modification in magnetic properties due to size reduction has several implications. Due to the small size of the particles a coherent rotation of the spins is possible (single domain nature). On the basis of such a scenario the bifurcation of the ZFC-FC magnetization can be attributed to the blocking of the magnetic moments of the nanoparticles. The broad peak in the ZFC curve is supposed to denote the average blocking temperature ($T_B$) of the system. Assuming that the dc measurement time scale equals ~100 s one can show that $25\kappa_B T_B \approx KV$, where K is the uniaxial anisotropy and V is the volume of the particles constituting the system, respectively [28]. Obviously, in the present case, due to the distribution of the particle size there will be a distribution of the blocking temperature in the samples. This is reflected in the fact that the bifurcation between the ZFC and FC magnetization curves starts at a higher temperature ($T_{irr}$) than the blocking temperature $T_B$ in all the three nanometric samples (S850, S750 and S600). The reason is that moment of some of the particles having bigger size starts blocking at higher temperatures than $T_B$. Moreover, the peak in the ZFC magnetization shifts towards the lower temperature with the decrease of particle size corroborating the direct relation of blocking temperature with the average volume/size of the particles.

Such nanoparticles with FM character often behave like superparamagnets above $T_B$. To verify whether the nanoparticles have superparamagnetic character, we plot M of S750 sample as a function of H/T at different temperatures in the regime $T_B<T<T_C$ (Fig. 6). According to Langevin's equation given by,

$$\frac{M}{M_0} = \coth\left(\frac{\mu_0 \mu H}{\kappa_B T}\right) - \frac{\kappa_B T}{\mu_0 \mu H} \tag{7}$$

where, $\mu$ is the moment of a particle and $M_0$ is the saturation magnetization, the magnetization plotted as a function of H/T at different T should overlap on a single curve. From Fig. 6 we find that the curves do not overlap and thus strongly suggests the absence of ideal superparamagnetic (SPM) behavior in the nanoparticles.

### 3.3. Linear and non-linear ac susceptibility study

To probe the magnetic state with more details the ac magnetic studies have been performed in terms of linear and non-linear ac magnetic susceptibility measurements. The magnetization M of a sample, in general, can be expressed as,

$$M = M_0 + \chi_1 H + \chi_2 H^2 + \chi_3 H^3 + ...,$$

(8)

where, $M_0$ is the spontaneous magnetization and H is the applied magnetic field. Figure 7 (a) shows the real part of linear susceptibility ($\chi_1^R$) of all the samples as a function of temperature. The magnitude of $\chi_1^R$ of S1150 sample is extremely small compared to that of the other samples as it (S1150) has antiferromagnetic character. Moreover, this sample shows a peak around $T_N$ similar to what is observed in the dc magnetization data. On reduction of particle size the peak of $\chi_1^R$ vs. T gets broadened and the magnitude of $\chi_1^R$ increases by order of magnitude. The peak is also found to shift to the lower temperature with the decrease of particle size. The measured real part of second harmonic ($\chi_2^R$) of the samples as a function of temperature are shown in the Fig. 7(b). All the samples display a broad peak around the transition temperatures. One can observe that the magnitude of $\chi_2^R$ has increased to a great extent on reduction of particle size compared to the bulk-like S1150 sample. As the origin of $\chi_2$ is the presence of spontaneous magnetization [29], these results basically reconfirms that the ferromagnetic character has been enhanced on decrease of particle size in our system. The $\chi_3^R$ vs. T curves for all the samples as shown in the Fig. 7(c) display similar behavior – a crossover from a positive (a broad positive peak) to a negative (a broad negative peak) value around the FM-PM transition temperature. This nature is observed generally in ferromagnets [30]. Our ac susceptibility data also indicates a size-induced FM state in the samples.

Further, we examine whether the nanoparticles follow the Wohlfarth's model of superparamagnetism. According to this model [31],

$$\chi_1 = n<\mu>/3\kappa_B T \qquad\qquad (9)$$

and

$$\chi_3 = - (n<\mu>/45)(<\mu>/\kappa_B T)^3 \quad . \qquad\qquad (10)$$

Here, n is the total number of magnetic clusters in the sample and $<\mu>$ is the average magnetic moment of a single cluster (particle). From this equation it is clear that above $T_B$ (in the SPM regime) $\chi_1$ is proportional to $T^{-1}$ and $\chi_3$ is proportional to $T^{-3}$. We plot $\chi_1^R$ vs. $T^{-1}$ and $\chi_3^R$ vs. $T^{-3}$ above $T_B$ for S750 samples in the Fig. 7 (d) and (e), respectively. Above $T_B$ the linearity is not appreciable. Moreover, only over a narrow temperature regime (around $T_C$) nearly linear behavior is observed. This again indicates that the nanoparticles with FM character do not behave like ideal superparamagnets.

### 3.4. Probing the inter-particle interaction

For magnetic nanoparticle systems one fundamental issue is the interaction among the particles which influences most of the magnetic properties of the systems. This interaction is also one of the reasons for not observing ideal superparamgnetic behavior in ferromagnetic nanoparticles. In our compacted system of nanoparticles, having a closed packed arrangement, presence of interaction between the particles seems to be highly probable. We have performed experiments on this issue to probe the inter-particle interaction in the nanometric samples (S750 and S600) through measurement of isothermal remanent magnetization and dc demagnetization as a function of magnetic fields [32].

Isothermal remanent magnetization (IRM) is the magnetization measured at a particular temperature after reducing an applied field ($H_{app}$) to zero isothermally. For measuring dc demagnetization (DCD) one has to apply a positive saturation field at particular temperature. Then a field in the opposite direction ($-H_{app}$) has to be applied. The measured magnetization after reducing $-H_{app}$ to zero is the DCD. The measured IRM and DCD (normalized) at 120 K for S750 sample are plotted in of Fig. 8 (a) as a function of $H_{app}$ (without sign) ranging from 0 to 500 Oe. For a non - interacting system with uniaxial anisotropy one can obtain the Wohlfarth relation [33]

$$m_{DCD} = 1 - 2\, m_{IRM}. \qquad (11)$$

Here 'm' is normalized magnetization with respect to the remanent magnetization (IRM) at $H_{app} = 500$ Oe. Due to the presence of any kind of interaction between the particle's magnetic moment, which either favours demagnetization or magnetization in the sample, this equality does not exist. So, the quantity

$$\delta m = m_{DCD} - (1 - 2\, m_{IRM}) \qquad\qquad (12)$$

is commonly used to probe the interaction strength between the particles [32]. A non- zero and negative value of $\delta M$ as shown in        Fig. 8 (b) for both the samples (S600 and S750) indicates that the particles are interacting and the interaction is of dipolar type, respectively. Negative $\delta m$ corresponds to the fact that this interaction actually favours the demagnetized state in the sample [32]. The (broad) minima of $\delta M$ vs. $H_{app}$ curves for S750 and S650 samples are found to be at around 200 Oe and 150 Oe, respectively. These values of $H_{app}$ are the respective average dipolar interaction strength of these samples at 120 K.

### 3.5. Magnetic relaxation behavior: aging and memory

Now, we attempt to investigate the magnetic relaxation behavior of the nanopaticles. To measure magnetization relaxation in the nanoparticles, the samples (S750 and S600) were cooled in a field of 100 Oe down to a measurement temperature of 120 K from above the blocking temperature and then we switched off the magnetic field immediately or after a waiting time $t_w$. Immediately after switching off the field, we measured the magnetization (known as thermoremanent magnetization) of the samples with respect to time (t). The magnetization of both the samples is found to display logarithmic behavior with time as shown in the Fig. 9 (a). These M vs. t curves have been fitted satisfactorily (for $t_w = 0$ s) with the equation

$$M = M_0 - S \log (t) \qquad\qquad (13)$$

where S is known as the magnetic viscosity (Insets of Fig. 9 (a)). Most interestingly, for S750 sample, the curve obtained after waiting time   $t_w = 3000$ s does not overlap with that for $t_w =$ 0 s indicating a waiting time dependence of the relaxation i.e. presence of aging phenomena. Moreover, both the S750 and S600 samples are found to display memory effect as shown in the Fig. 9 (b) and in its inset, resepectively. To perform the test for memory effect for S750 sample, we cooled the sample in a field of 100 Oe and simultaneously magnetization was measured (FCC_mem) with an intermediate halt at 120 K for 3 h while cooling. During this time of halt field was remained switched off. We then resumed the cooling and magnetization measurement with same applied field down to the lowest temperature. Magnetization was again measured in the warming cycle (FCW_mem) with same field applied. Interestingly, we find that the sample remembers the measurement history as follows. While warming, FCW_mem data follows the FCC_mem data and then displays a jump at around 120 K (at this temperature field was switched off during cooling) and follow the FCC_mem data again (Fig. 9 (b)). For comparison, the reference curve is also shown (FCW_ref) which is measured magnetization during warming without intermediate switching off the field during field cooling. A similar memory effect has also been found in the S600 sample as shown in the inset of Fig. 9 (b). In this case we switched off the field at 95 K and 120 K for 2 h at each temperature. While warming, we again observe a prominent jump of the FCW_mem magnetization curve at 95 K. However, very weak anomaly is found around 120 K. These results are precisely an exhibition of memory effect. Such aging and memory effect are generally observed in spin glasses. It has been found earlier that in a system of magnetic nanoparticles, the correlation between the particles develops in such a way that the system displays many spin glass like behaviors [34,35].

Two theoretical models, namely, droplet [36] and hierarchical [37], are very often used to explain the presence of memory effect and other issues in spin glasses. According to droplet model a positive and negative temperature change (by same amount) affects the magnetic relaxation behavior symmetrically. According to hierarchical model, on the other hand, the behaviors should be asymmetric with positive and negative temperature change. We now perform the magnetization relaxation experiment with positive and negative temperature cycle. After the sample (S750) is zero-field-cooled to a measurement temperature of 120 K, a magnetic field of 100 Oe is switched on and the magnetization is measured for a period of time $t_1$. Then the sample is cooled down to a temperature of 100 K without changing the field and measurement is resumed and performed for a period of time $t_2$. Finally, the sample is heated back to 120 K and the magnetization is measured with time for a period $t_3$. The resulting M vs. t curves are shown in the Fig. 10 (a). We note that the relaxation of magnetization is almost stopped at 100 K showing a constant magnitude of M with $t_2$. Interestingly, M starts changing (relaxation process restarts) when the system is heated back to 120 K. A closer observation reveals that the relaxation of magnetization during $t_3$ is nearly a continuation of the relaxation during $t_1$. This measurement was repeated with only an intermediate heating of the sample to 140 K instead of cooling and shown in the Fig. 10 (b). In this case, we see that the magnetization relaxes during $t_1$ and $t_2$ but relaxation is almost halted during $t_3$. Clearly, this behavior on positive temperature change is different from those observed for negative temperature change. This asymmetric behavior in magnetization relaxation with respect to temperature change in our samples indicates that the system dynamics has similarity as described under hierarchical model rather than droplet model. To further confirm our conjecture we perform the same experiment in case of the decay of thermoremanent magnetization for S600 sample. The sample was field-cooled    (100 Oe) down to   85 K and the magnetization was measured with time after switching off the field. Then we applied an intermediate cooling and heating by the same amount (20 K) and continued the relaxation measurements. The resulted curves are shown in the insets of Fig. 10 (a) and (b). We observe that after cooling the change in M is halted while the relaxation resumes on heating to the initial temperature (Inset of Fig. 10 (a)).  In contrast, after the intermediate heating M keeps on relaxing while the relaxation is stopped on cooling back to the initial temperature (Inset of Fig. 10 (b)). These results are exactly same to the observed effects of temperature cycling on magnetization relaxation of S750 sample. According to the hierarchical model, as per a qualitative description, multivalley structure is organized on the free-energy surface at a given temperature. The free-energy valleys (metastable states) split into new sub valleys with decreasing temperature. Relaxation occurs only within the valleys. On re-heating the newly developed sub valleys and barriers merge back to the previous free-energy landscape. Relaxation is then restarted. Hence, our results imply that a hierarchical organisation of metastable states is present in these nanoparticle systems. Large number of interacting particles also fulfils the criteria of having large number of degrees of freedom in hierarchical organization.

### 3.6. Field dependence of the blocking temperature

We know that the field dependence of freezing temperature of spin glasses has been studied with some details earlier. It has been found that on de Almeida-Thouless (AT) line the freezing temperature is proportional to $H^{2/3}$ and mostly followed by the anisotropic Ising spin glasses [38]. On the other hand, on Gabay-Toulouse (GT) line the freezing temperature is proportional to $H^2$ as found in case of isotropic Heisenberg spin glasses [39]. However, it has also been found that in the systems with SPM nature and existing inter-particle interaction these laws can describe the field dependence of the blocking temperature as well [40]. We have investigated this issue by measuring $T_B$ at different fields in the nanoparticles. Figure 11 shows the variation of $T_B$ as a function of $H^{2/3}$ (main panel) and $H^2$ (in the inset) for S750 and S600 samples. It is clearly observed that these samples follow the AT law and deviates from it slightly at higher magnetic field. The GT law is not found to hold at al. Earlier, a similar observation has been reported in nanoparticles of $Pr_{0.5}Sr_{0.5}MnO_3$, where a cross over from a AT to GT line was found [40]. We find from Fig. 11 (a) that the deviation from linearity starts around 150 -200 Oe for the samples. This value of the field is, on the other hand, nearly equal to the dipolar interaction strength for the samples. This observation can be related to the suppression of the anisotropy (due to the dipolar interaction) by the magnetic field.

## 4. Discussions

We, now critically analyze the observed magnetic behavior of the samples.

(i) As we have mentioned earlier that the bulk sample of $Nd_{0.4}Sr_{0.6}MnO_3$ has an A-type antiferromagentic ground state. In this type of spin ordering the ferromagnetically ordered layers (all the spins are oriented parallel in one layer) are stacked antiferromagnetically. Obviously, the net moment becomes zero. In such a magnetic state, one can easily understand that a slight destabilization will result in a non-zero net moment or a partial ferromagnetism as ferromagnetic order already exists. We know that the nanoparticles are associated with surface disorder. The surface consists of various defects, vacancies broken bonds etc. So naturally the magnetic order is predominantly disturbed in the surface region and uncompensated spins will produce some net moment. As the size of the particle is reduced the surface to volume ratio is increased and enhances the FM character in the system. The resulting scenario is well described in terms of a core-shell type model. The core is magnetically ordered (AFM type), whereas, the shell becomes partially ferromagnetic.

Such surface disorder can be a natural ingredient for the observed GP in the nanoparticles. Although the existence of Griffiths like phase has been observed in other nanometric manganites previously [21,22], no comprehensive explanation for the occurrence of this phase is available till date. Since one precursor for Griffths like phase is disorder [17,21,22], the enhanced disorder effect due to size reduction can be related to the observation of such a phase in the nanometric system. The disorder presumably persists around the surface region where there is a possibility of random spatial distribution of the exchange interaction which renders the Griffiths like phase.

(ii) We have found earlier that the estimated critical exponents do not fall in any of the common universality classes (Heisenberg, Ising or mean field). This may be due to the presence of some amount of AFM phase in the sample and unusual nature of the size-induced FM state. We also find that the Arrott plots (Fig. 4(a)) for S750 sample have a peculiar nature showing an upward concave curvature unlike that for ordinary ferromagnets where a downward concave curvature is generally observed. This again indicates that the magnetic state of the S750 sample is different from an ordinary ferromagnetic manganite. In some recent studies it has been found that the finite size of a system can hinder the divergence of coherence length [41]. Such effects are very likely to be present in our system nanometric system (S750) and may lead to the unusual values of the critical exponents.

(iii) Employing Rhodes-Wohlfarth [42] criterion for S750 sample we find that $q_c/q_s > 1$ which indicates the itinerant character of the induced FM state in S750 sample. Here, $\mu_{eff} = \sqrt{q_C(q_C+2)}$ and $q_S$ is the saturation moment of the sample. We have found that $q_C = 3.92$ and $q_S = 2.04$. This itinerant character could be another reason for getting different values of the critical exponents from those determined on the basis of standard localized models.

(iv) Interestingly, we observe that the value of $T_C$ for S750 sample is higher than the value of $\theta_{CW}$. Generally, $\theta_{CW}$ is found to be equal to $T_C$ or higher than that (due to the presence of spin clusters [43]). The reduction of $\theta_{CW}$ even below $T_C$ in case of S750 sample can be attributed to the presence of AFM phase in the sample. For antiferromagnetic materials $\theta_{CW}$ is negative. So, presence of FM and AFM phase in this sample leads to a value of $\theta_{CW}$ which is below the FM-PM transition temperature ($T_C$).

(v) The observed slow logarithmic type relaxation of magnetization, aging as well as the memory effect can be attributed to the broad distribution of relaxation time along with the existing inter-particle interaction [44,45]. The relaxation time for a particle having uniaxial anisotropy K and volume V is given by [28]

$$\tau = \tau_0 \exp{(KV/\kappa_B T)} \ . \qquad (14)$$

One can easily understand that the distribution of particle size (V) as evident from Fig. 1 (d) and (f), naturally gives rise to a broad distribution of relaxation time (including very large $\tau$) due to the exponential nature of the function. We try to explain the observed memory effect qualitatively on the basis of distribution of the relaxation time. There is a fraction of small particles whose relaxation time is smaller than the experimental time scale ($\tau_e$) even below the memory measurement temperature ($T_m$) where the field was switched off. These particles will always be in the equilibrium superparamagnetic state and will display Curie-type behavior ($M \propto 1/T$) during heating as well as cooling. These particles are not of much interest. However, the fraction of bigger particles having $\tau > \tau_e$ below $T_m$ or even above this temperature play significant role in the memory effect. Some of these bigger particles block towards their random anisotropy direction when the field is remained switched off at $T_m$ for long time such that even resuming the magnetic field magnetization does not come back to the previous value within the observation time showing a gap between FCC_mem and FCW_ref curves below $T_m$ (Fig. 9 (b)). During warming cycle, magnetization (FCW_mem) increases as $T_m$ is approached and on further increment of temperature more number of these bigger particles (those blocked just below $T_B$) enters the superparamagnetic regime. When the M vs. T curves of these smaller and bigger particles are superimposed it simply looks like the FCW_mem curve as shown in the Fig. 9 (b). Hence, we see that even a broad distribution of particle size (and $\tau$) can lead to memory effect. However, due to the presence of interaction between the nanoparticles the relaxation time is modified and the collective dynamics of the particles are influenced [43]. So the present observation of memory effect is attributed to the combined effect of broad distribution relaxation time and inter-particle interaction.

## 5. Conclusions

In summary, we can say that on reduction of size AFM state is destabilized in $Nd_{0.4}Sr_{0.6}MnO_3$. Ferromagnetic ground state is found in the nanoparticles of this system. Existence of a Griffiths like phase is observed in the nanoparticles. This size-induced FM state is different from an ordinary ferromagnet as evident from the significant deviation of the critical exponents estimated from modified Arrott plot (and KF plot) from those predicted under standard models. Dipolar type interaction is found in the nanoparticles and ideal superparamagnetism is not observed in their magnetic behavior. The nanoparticles display logarithmic type magnetization relaxation, aging and memory effect. Such spin glass like properties (i.e. aging and memory) are attributed to broad distribution of relaxation time and inter-particle interaction. The blocking temperature of the nanometric samples follows a de Almeida-Thouless type field dependence in the low field regime.

**Acknowledgement**

One of the authors (T. K. Nath) would like to acknowledge the financial assistance of Department of Science and Technology (DST), New Delhi, India through project no. IR/S2/PU-04/2006.

**Figure caption**

**Figure 1.** (Colour online). (a) The experimental x- ray diffraction data and fitted curve (computed data) after Rietveld refinement of (a) S1150 sample, (b) S600 sample. (c) TEM image of the particles of S600 sample. (d) The size distribution of particles of S600 sample. (e) TEM image of the particles of S750 sample. (f) The distribution of particle size of S750 sample. (g) FESEM image of S850 sample. (h) FESEM image of S1150 sample.

**Figure 2.** (Colour online). The zero-field-cooled (ZFC) and field - cooled (FC) magnetization of S600, S750 and S850 samples measured at 50 Oe. Inset shows the same for S1150 sample measured at 350 Oe. Larger symbols in the plots represent the FC curves and the smaller symbols represent the ZFC curves.

**Figure 3.** (a) (Colour online). M-H isotherm of the S750 sample measured at 10 K. The inset of (a) shows the Arrott plot at the same temperature. (b) Plot of $1/\chi$ of this sample as a function of T in the high temperature paramagnetic regime. The line is the fitted straight line employing Curie-Weiss law. (c) The plot of $\chi^{-1}$ as a function $(T/T_C^R-1)$ in log scale for S750 sample. The lines are the fitted curves (straight lines) in linear portions of the graph.

**Figure 4.** (Colour online). (a) Arrott plot at different temperatures for S750 sample. The curves display an upward concave curvature. (b) Modified Arrott plot ($M^{1/\beta}$ vs. $(H/M)^{1/\gamma}$) around critical regime of this sample.

**Figure 5.** (Colour online). (a) Kouvel-Fisher plot with linear fits for the S750 sample. The inset displays the critical isotherm in log-log scale with the fitted straight line. (b) The plot of $M|\varepsilon|^{-\beta}$ vs. $H|\varepsilon|^{-(\gamma+\beta)}$ in log-log scale showing two different branches ( for $T<T_C$ and $T>T_C$) and validity of the scaling theory.

**Figure 6.** (Colour online). The plot of magnetization (M) as a function of H/T at different temperatures in the regime $T_B < T < T_C$ for S750 sample.

**Figure 7.** (Colour online). The plot of (a) $\chi_1^R$, (b) $\chi_2^R$, and (c) $\chi_3^R$ of all the samples as a function of temperature. The plot of (d) $\chi_1^R$ as a function of $T^{-1}$ and (e) $\chi_3^R$ as a function of $T^{-3}$ for S750 sample.

**Figure 8.** (Colour online). (a) The plot of normalized IRM and DCD magnetizations of S750 sample as a function of applied field $H_{app}$ (The sign of $H_{app}$ has been ignored). (b) Variation of $\delta m$ with $H_{app}$ for S750 and S600 samples.

**Figure 9.** (Colour online). (a) Variation of normalized thermoremanent magnetization (H=100 Oe) of S750 sample as a function of time for different waiting times showing the aging effect at 120 K. The left inset displays the same for $t_w$= 0 s for S750 sample in logarithmic time scale showing linear behavior (M=$M_0$-Slog(t)) as evident from the fitting with straight line. The right inset shows the decay of thermoremanent magnetization along with the fit employing the same equation for S600 sample at 120 K. (b) The measured magnetization as a function of temperature display the memory effect in S750 as described in the text. The inset shows the similar memory effect for S600 sample.

**Figure 10.** (Colour online). (a) Effect of negative temperature change on the relaxation of magnetization for S750 sample. The details regarding this measurement are given in the text. The inset shows the effect of negative temperature change on the decay of thermoremanent magnetization of S600 sample. (b) Similar measurements as in (a) for S750 and S600 samples with a positive temperature change.

**Figure 11.** (Colour online). (a) Variation of blocking temperature ($T_B$) with $T^{2/3}$ (main panel) and with $T^{1/2}$ (inset) for S750 sample. (b) Variation of blocking temperature ($T_B$) with $T^{2/3}$ (main panel) and with $T^{1/2}$ (inset) for S600 sample. The lines are guide to the eye.

**TABLE 1**

Various estimated parameters related to crystal-structure (refined parameters), morphology and magnetic properties of S600, S750, S850 and S1150 samples. Possible error in the values is shown in the brackets wherever significant.

| Sample Name | S600 | S750 | S850 | S1150 |
|---|---|---|---|---|
| space group | *I4/mcm* | *I4/mcm* | *I4/mcm* | *I4/mcm* |
| a (Å) | 5.4144 (3) | 5.4057(4) | 5.4382(4) | 5.3854(2) |
| c (Å) | 7.7161(9) | 7.6989(10) | 7.6331(9) | 7.7822(3) |
| V (Å$^3$) | 226.2 | 225.0 | 225.7 | 225.7 |
| Crystallite size (nm) (from HRXRD) | 20 | 29 | 42(1) | 217(3) |
| Particle size (nm) (from HRTEM and FESEM) | 27 (7) | 45 (10) | 95 (15) | 300 (50) |
| T$_B$ (K) (at 50 Oe) | 160 (5) | 190 (4) | 200 (5) | --- |

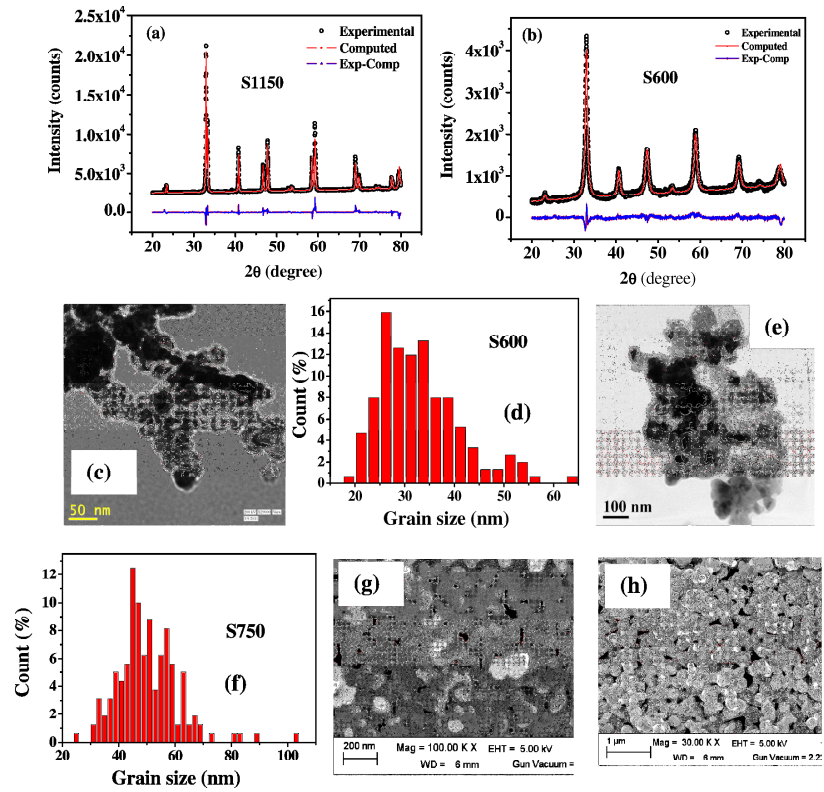

**Fig. 1**

**S. Kundu et al.**

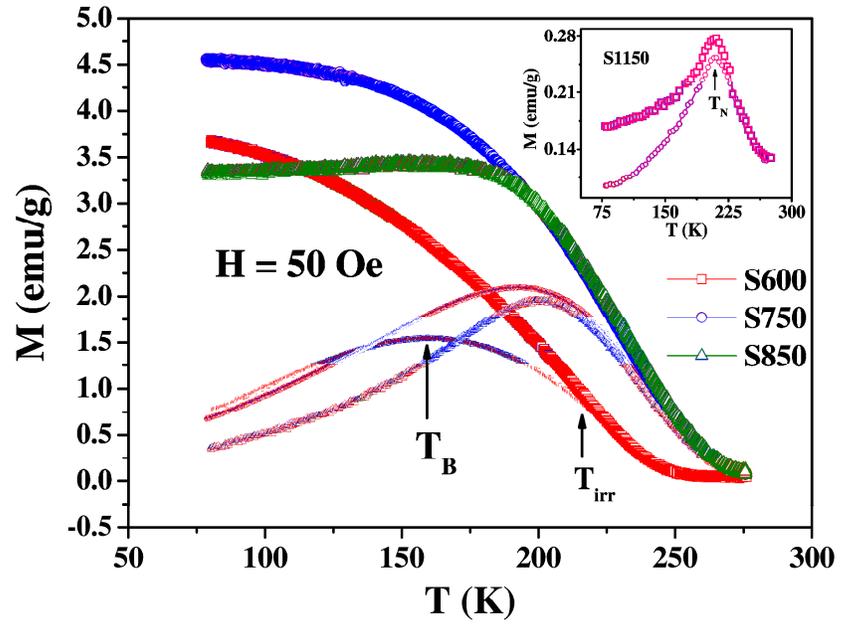

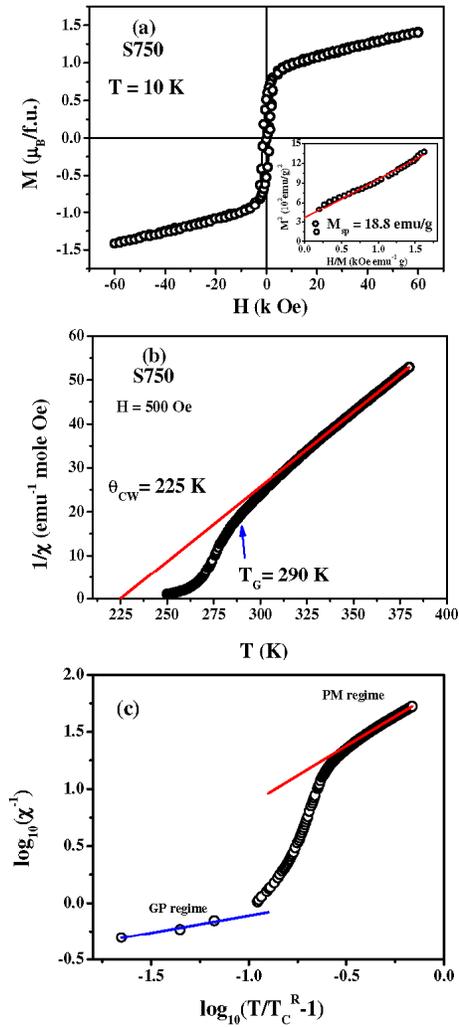

Fig. 3

S. Kundu et al.

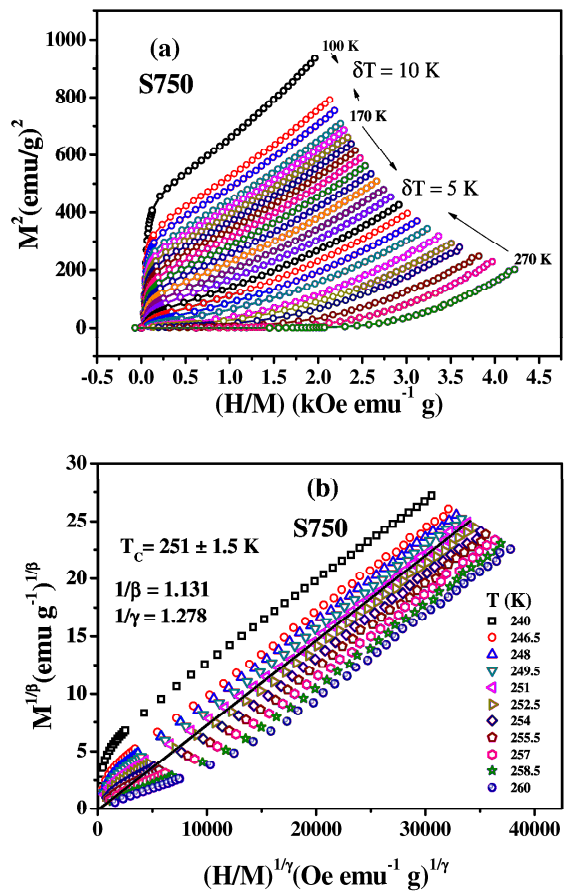

**Fig. 4**

**S. Kundu et al.**

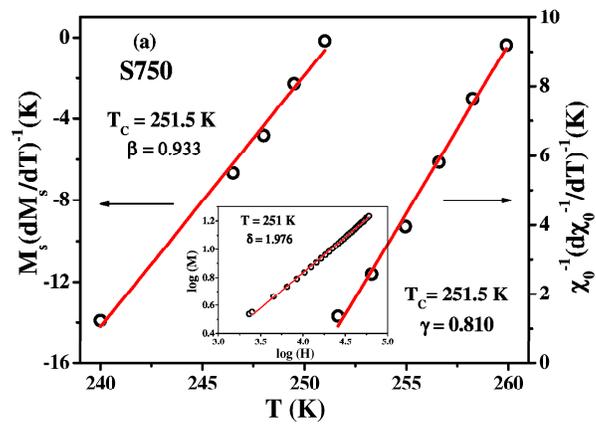

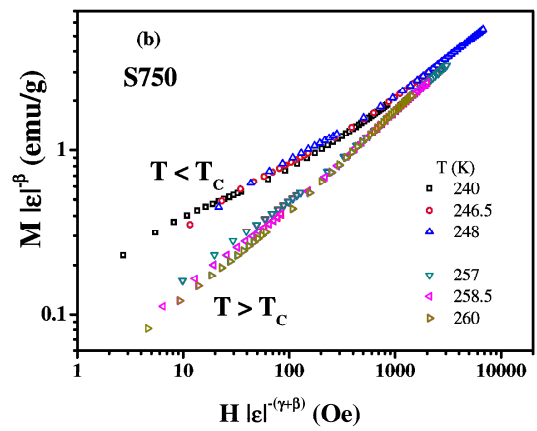

Fig. 5

S. Kundu et al.

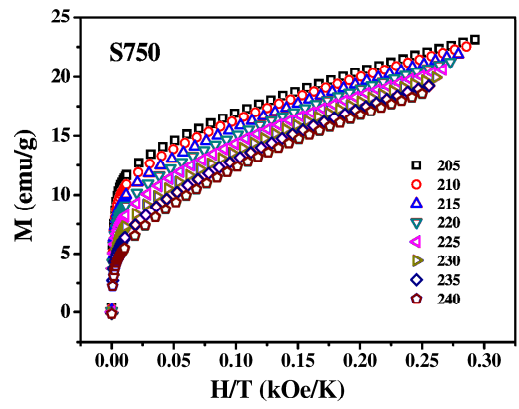

**Fig. 6**

**S. Kundu et al.**

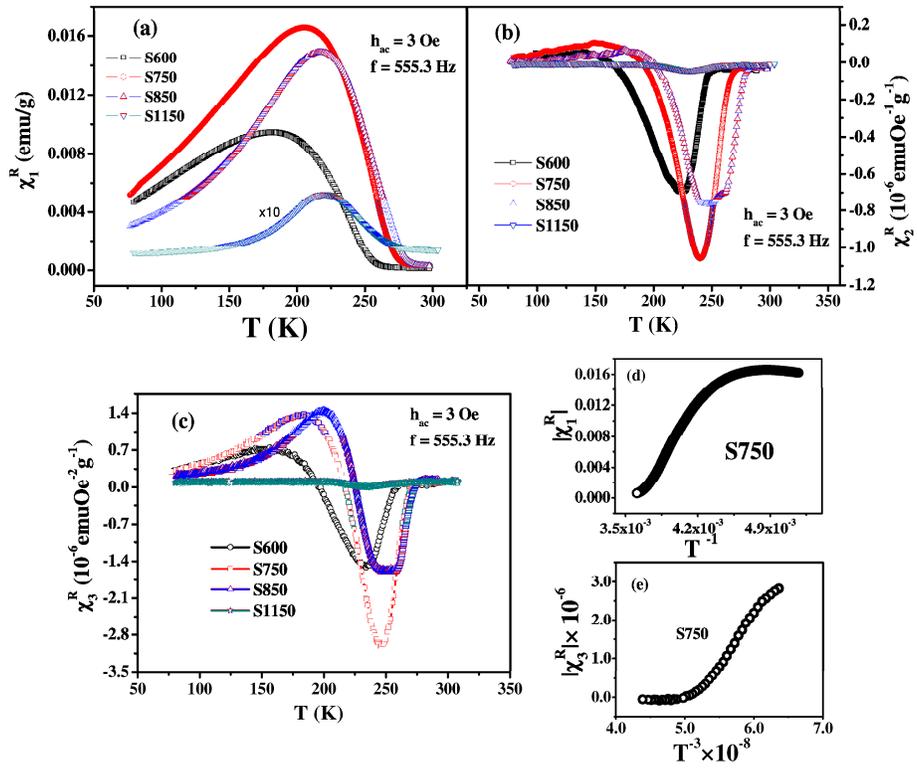

**Fig. 7**

**S. Kundu et al.**

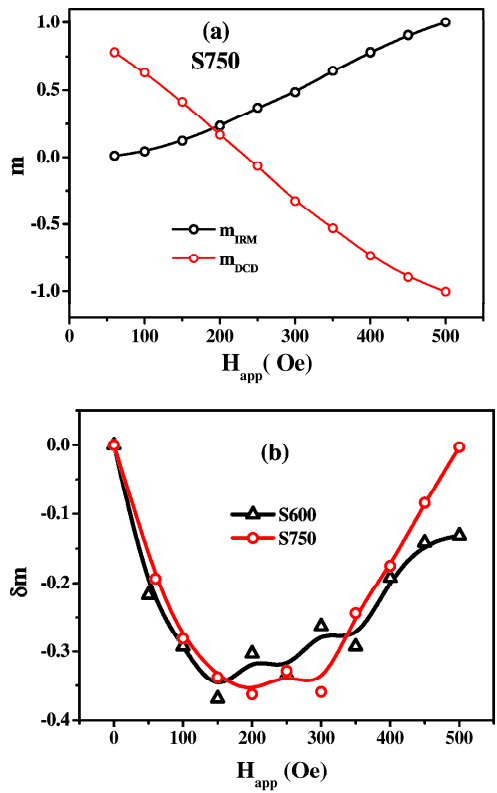

**Fig. 8**

**S. Kundu et al.**

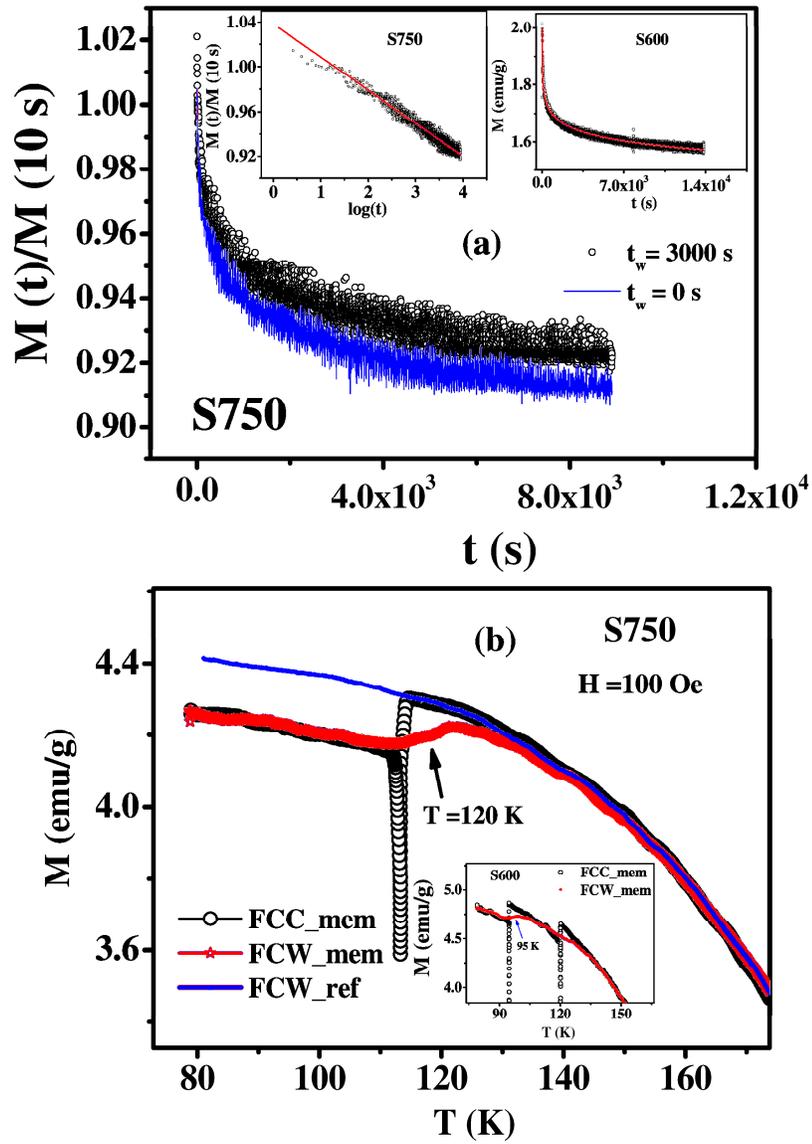

**Fig. 9**

**S. Kundu et al.**

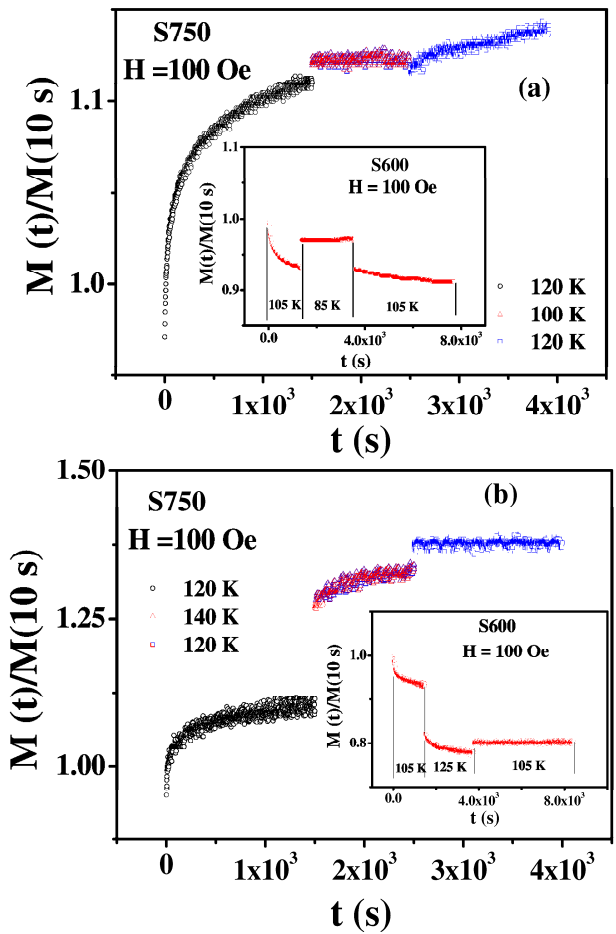

**Fig. 10**

**S. Kundu et al.**

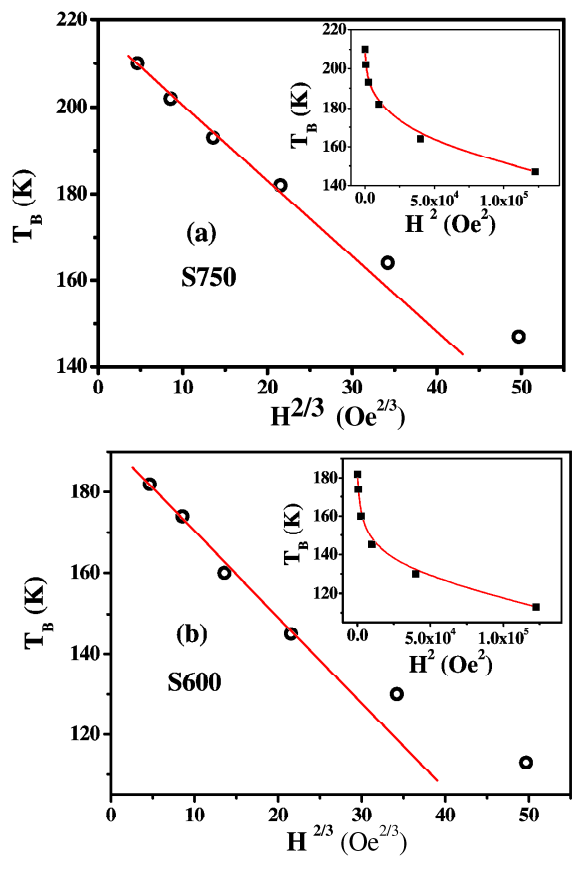